\documentclass[12pt]{article}
\usepackage{amsfonts}
\usepackage{latexsym}
\usepackage{amsmath}
\usepackage{amssymb}
\usepackage{amssymb}
\usepackage{slashed}
\usepackage{xcolor}

\renewcommand{\thefootnote}{\arabic{footnote}}

\renewcommand{\thefootnote}{\fnsymbol{footnote}}
\newcommand{\newsection}{    
\setcounter{equation}{0}\section}
\def\appendix#1{\addtocounter{section}{1}\setcounter{equation}{0}
\renewcommand{\thesection}{\Alph{section}}
\section*{Appendix \thesection\protect\indent \parbox[t]{11.15cm}{#1}}
\addcontentsline{toc}{section}{Appendix \thesection\ \ \ #1}}

\renewcommand{\thefootnote}{\arabic{footnote}}

\begin{document}

\begin{titlepage}
\begin{center}

\vspace*{-1.0cm}

\hfill  DMUS-MP-21/09
\\
\vspace{2.0cm}

\renewcommand{\thefootnote}{\fnsymbol{footnote}}
{\Large{\bf Supersymmetry Enhancement of Heterotic Horizons}}
\vskip1cm
\vskip 1.3cm
D. Farotti and J. Gutowski
\vskip 1cm
{\small{\it
Department of Mathematics,
University of Surrey \\
Guildford, GU2 7XH, UK.}\\
\texttt{d.farotti@surrey.ac.uk, j.gutowski@surrey.ac.uk}}

\end{center}

\bigskip

\begin{abstract}

\end{abstract}

The supersymmetry of near-horizon geometries in heterotic supergravity is considered.
A necessary and sufficient condition for a solution to preserve more than the minimal
$N=2$ supersymmetry is obtained. A supersymmetric near-horizon solution is constructed which is a $U(1)$ fibration of $AdS_3$ over a particular Aloff-Wallach space. It is proven that this solution preserves the conditions required for $N=2$ supersymmetry, but does not satisfy the necessary condition required for further supersymmetry enhancement. Hence, there exist supersymmetric near-horizon heterotic solutions preserving exactly $N=2$ supersymmetry.

\end{titlepage}



\setcounter{section}{0}
\setcounter{subsection}{0}
\setcounter{equation}{0}

\newsection{Introduction}

The geometric properties of horizons of supersymmetric black holes are very closely linked to 
the notion of supersymmetry enhancement. In particular, the presence of additional spinors ensures that the black hole near-horizon solutions have certain symmetries. The first well-understood example of this 
is the case of the BMPV black hole \cite{Breckenridge:1996is}. The bulk geometry of
this solution preserves half of the supersymmetry, whereas the near-horizon solution
obtained by taking an appropriate decoupling limit is maximally supersymmetric. 
Indeed, the systematic analysis of the near-horizon geometries of minimal $N=2, D=5$
supergravity constructed in \cite{Reall:2002bh} found all possible near-horizon solutions
of this theory, including flat space and $AdS_3 \times S^2$, which are also maximally supersymmetric. By exploiting certain similarities between heterotic supergravity 
and $N=2$, $D=5$ supergravity, this near-horizon supersymmetry doubling was also proven
to hold for heterotic horizons in \cite{hhor1}, utilizing the classification of supersymmetric heterotic solutions in \cite{Gran:2005wf, Gran:2007kh}. Consequently, the number of
supersymmetries of heterotic near-horizon solutions was shown to be even.
Further investigation of near-horizon geometries, utilizing generalized Lichnerowicz theorems, extended this result to $D=11$ supergravity \cite{Gutowski:2013kma},
and also to type II supergravities in $D=10$ \cite{Gran:2013wca, Gran:2014fsa, Gran:2014yqa}.
In particular, as a consequence of this, black hole near-horizon geometries generically admit a
$SL(2, \mathbb{R})$ symmetry algebra. 

The conditions for the heterotic horizons are notably similar to those of the 
$N=2, D=5$ horizons. They are also rather stronger
than those found for near-horizon geometries in type II and $D=11$ supergravity,
for both $N=2$ and $N \geq 4$ supersymmetry. In particular, the conditions
required for near-horizon geometries to preserve $N=4$ supersymmetry in $D=11$
supergravity were found in \cite{Farotti:2021otm}. As the structure of the heterotic 
near-horizon solutions are somewhat simpler, it is most straightforward to find explicit examples
of such solutions in the heterotic theory. Some progress was made in \cite{hhor2}, where a large class of heterotic
horizons were found based on del Pezzo surfaces. However, all of these examples preserve
at least $N=4$ supersymmetry, and following \cite{hhor1}, it is known that
the minimal amount of supersymmetry for heterotic near-horizons is $N=2$.

The purpose of
this work is to prove that there do exist solutions preserving exactly $N=2$ supersymmetry.
As has been proven in \cite{hhor1}, the number of supersymmetries of such solutions
must be even, so there can be no $N=3$ solutions. However, it has been unclear if there
may be some mechanism whereby supersymmetry might be automatically enhanced from
$N=2$ to $N=4$, because hitherto there have been no explicit exactly $N=2$ solutions.
We shall establish the existence of exactly $N=2$ solutions first by
establishing a particularly simple condition which is necessary
and sufficient for a $N=2$ solution to preserve additional $N \geq 4$ supersymmetry. Then we
proceed to construct an explicit near-horizon geometry, which is a $U(1)$ fibration of
$AdS_3$ over a certain 7-dimensional Aloff-Wallach space $M_{k,\ell}=SU(3)/U(1)_{k, \ell}$ equipped with co-closed $G_2$ structure.
This solution satisfies the conditions required for $N=2$ supersymmetry, but fails
to satisfy the condition which is necessary and sufficient for further supersymmetry enhancement.
In particular, this solution is an example of a ``descendant" solution,
for which the gravitino equation holds for all the spinors, but the dilatino equation does
not hold for two of the spinors
\cite{Papadopoulos:2009br, Gran:2007fu}. This establishes the existence of exactly $N=2$ supersymmetric heterotic near-horizon geometries.

The plan of this paper is as follows. In Section 2 we summarize some of the key details
concerning supersymmetric heterotic near-horizon geometries derived in \cite{hhor1}.
In Section 3 we demonstrate how the bosonic field equations, Bianchi identities and
Killing spinor equations reduce to conditions on a 7-dimensional manifold equipped
with a conformally balanced $G_2$ structure, and we also establish a Lichnerowicz type
theorem for the near-horizon geometries. In Section 4, utilizing the Lichnerowicz
type theorem, we construct a necessary and sufficient condition for a $N=2$ supersymmetric
heterotic near-horizon solution to admit $N \geq 4$ supersymmetry. In Section 5, an explicit
construction of an exactly $N=2$ near-horizon geometry is provided, utilizing a $G_2$ structure
defined on a certain Aloff-Wallach space. Section 6 contains some conclusions. In Appendix A,
some useful $G_2$ identities are listed. In Appendix B, some further details relating
to the Aloff-Wallach space used to construct the solution in Section 5 are given.
In Appendix C, properties of the Fern\'andez-Gray \cite{fgclass} classification of $G_2$ structures
are listed.

\newsection{Supersymmetric Heterotic Horizons}

In this section we summarize some particularly important results
for supersymmetric heterotic near-horizon geometries,
as found in \cite{hhor1}.
These results were found by using a bilinear matching
condition to simplify some of the bosonic fields in the solution. However, as has been 
shown in \cite{Gran:2013wca},
these conditions can be obtained independently via
a compactness argument.

The $D=10$ heterotic gravitino and dilatino Killing spinor equations (KSE) are given by
\begin{eqnarray}
\label{htkse1}
\nabla_\mu \epsilon - {1 \over 8} H_{\mu \nu_1 \nu_2} \Gamma^{\nu_1 \nu_2} \epsilon
= {\cal{O}}(\alpha'^2)
\end{eqnarray}
\begin{eqnarray}
\label{htkse2}
\big(\Gamma^\mu \nabla_\mu \Phi - {1 \over 12} H_{\nu_1 \nu_2 \nu_3}
\Gamma^{\nu_1 \nu_2 \nu_3} \big) \epsilon &=& {\cal{O}}(\alpha'^2)
\end{eqnarray}
where $\mu, \nu$ denote $D=10$ frame indices and in ({\ref{htkse1}}), $\nabla$ denotes the $D=10$ Levi-Civita connection; $H$ is the 3-form and $\Phi$ is the dilaton. The gaugino KSE is given by
\begin{eqnarray}
\label{htkse3}
\Gamma^{\mu \nu} F_{\mu \nu} \epsilon = {\cal{O}}(\alpha') \ .
\end{eqnarray}
To zeroth order in $\alpha'$ the conditions involving $F$ decouple completely from the remaining equations, and consequently in this work we are counting the number of solutions of ({\ref{htkse1}}) and ({\ref{htkse2}}), taking $F=0$.

We shall be considering supersymmetric solutions which are near-horizon geometries.
In what follows, we assume that the 8-dimensional spatial cross section of the event horizon 
${\cal{S}}$ is smooth and compact without boundary. 
Following \cite{hhor1}, the near-horizon metric can be written in 
Gaussian Null co-ordinates \cite{isen, gnull} $\{u, r, y^I \}$ where
$y^I$ are local co-ordinates on ${\cal{S}}$. The metric, in the near-horizon limit, is then
\begin{eqnarray}
ds^2 =2 du (du+r h) + ds^2({\cal{S}})
\end{eqnarray}
where $h$ is a ($u,r$-independent) 1-form on ${\cal{S}}$, and the metric on ${\cal{S}}$ also does not depend on $u,r$. 

We remark that the analysis of the supersymmetric near-horizon solutions when $\alpha' \neq 0$ has been done in 
\cite{Fontanella:2016aok}. It is notable that if $\alpha' \neq 0$, the gaugino equation ({\ref{htkse3}}) follows from 
({\ref{htkse1}}) and ({\ref{htkse2}}), together with the bosonic field equations and Bianchi identities.
If $\alpha' \neq 0$ then the supersymmetric near-horizon solutions split into two cases. Firstly, if $h$ is covariantly constant, to zero and first order in $\alpha'$,
with respect to the metric connection on ${\cal{S}}$ whose torsion is equal to the pull-back of
$H$ to ${\cal{S}}$, then the number of supersymmetries is even (again, to zero and first order in $\alpha'$).  However, there is also a case for which this covariant constancy condition on $h$ does not hold to first order in $\alpha'$, and then the number of supersymmetries need not be even. It would be interesting to understand this case better. However, from 
\cite{hhor1} it is known that the covariant constancy condition on $h$ must hold to zeroth order in $\alpha'$ and consequently at zeroth order in $\alpha'$, the number of supersymmetries is even.

In addition to considering the metric in the near-horizon limit, we assume
that the 3-form $H$ admits a well-defined near-horizon limit, as 
considered in \cite{hhor1}. To zeroth order in $\alpha'$, $H$ is closed, with
\begin{eqnarray}
H=du \wedge dr \wedge h + r du \wedge dh + {\tilde{H}}
\end{eqnarray} 
where ${\tilde{H}}$ is a ($u,r$-independent) closed 3-form on ${\cal{S}}$.
Additional conditions are then obtained from the analysis of the
Killing spinor equations in \cite{hhor1}. These are:
\begin{eqnarray}
\nabla_{(i} h_{j)}=0
\end{eqnarray}
\begin{eqnarray}
{\cal{L}}_h \Phi =0, \qquad {\cal{L}}_h {\tilde{H}} =0
\end{eqnarray}
\begin{eqnarray}
h^2 = {\rm const.}
\end{eqnarray}
\begin{eqnarray}
dh = i_h {\tilde{H}}
\end{eqnarray}
\begin{eqnarray}
\label{gaugeeqn}
\nabla^k {\tilde{H}}_{kij} = 2 \nabla^k \Phi {\tilde{H}}_{kij}
\end{eqnarray}
\begin{eqnarray}
\label{deqn}
\nabla^2 \Phi = 2 \nabla^i \Phi \nabla_i \Phi +{1 \over 2}
h^2 -{1 \over 12} {\tilde{H}}_{ijk} {\tilde{H}}^{ijk}
\end{eqnarray}
\begin{eqnarray}
\label{eineq}
R_{ij} = {1 \over 4} {\tilde{H}}_{imn} {\tilde{H}}_j{}^{mn} -2 \nabla_i \nabla_j \Phi
\end{eqnarray}
where $i,j$ are frame indices on ${\cal{S}}$,
$\nabla$ denotes the Levi-Civita connection on ${\cal{S}}$,
and $R_{ij}$ is the Ricci tensor of ${\cal{S}}$. If $h=0$ then it has been shown that ${\tilde{H}}=0$ and $\Phi = {\rm const.}$, and the near-horizon geometry is $\mathbb{R}^{1,1} \times {\cal{S}}$ where ${\cal{S}}$ is a $Spin(7)$ holonomy manifold. We discard this special case.

In terms of explicitly counting the number of supersymmetries, it is useful to consider some algebraic properties of the Killing spinors, following the analysis of the KSE given in \cite{hhor1}. In particular, the Killing spinor $\epsilon$ of heterotic near-horizon solutions is determined algebraically
in terms of two spinors $\eta_{\pm}$ on ${\cal{S}}$ 
via 
\begin{eqnarray}
\epsilon = {u \over 2} h_i \Gamma^{-i}
\eta_- + \eta_+ + \eta_- \ ,
\end{eqnarray}
where
\begin{eqnarray}
\Gamma_{\pm} \eta_{\pm} = 0
\end{eqnarray}
and $+,-$ are lightcone directions associated with the 
Gaussian null co-ordinate system. The $D=10$ KSE decompose into conditions involving only
$\eta_+$, and conditions involving only $\eta_-$. Moreover, if
$\eta_+$ satisfies the KSE involving $\eta_+$, then
$\eta_-$ defined by
\begin{eqnarray}
\eta_- = \Gamma_- h_i \Gamma^i \eta_+
\end{eqnarray}
automatically satisfies the KSE involving $\eta_-$. 
Conversely, if
$\eta_-$ satisfies the KSE involving $\eta_-$, then
$\eta_+$ defined by
\begin{eqnarray}
\eta_+ = \Gamma_+ h_i \Gamma^i \eta_-
\end{eqnarray}
automatically satisfies the KSE involving $\eta_+$. It follows
that the total number of killing spinors $N$ must be even, $N=2N_+$,
where $N_+$ is equal to the number of positive chirality
spinors $\eta_+$. 

So, in analyzing the KSE, it suffices to consider the
KSE involving only $\eta_+$, which are given by
\begin{eqnarray}
\label{kse1}
\nabla_i \eta_+ = {1 \over 8} {\tilde{H}}_{ijk} \Gamma^{jk} \eta_+
\end{eqnarray}
\begin{eqnarray}
\label{kse2}
dh_{ij} \Gamma^{ij} \eta_+ =0
\end{eqnarray}
\begin{eqnarray}
\label{kse3}
(2 \nabla_i \Phi + h_i)\Gamma^i \eta_+ -{1 \over 6}
{\tilde{H}}_{ijk} \Gamma^{ijk} \eta_+ =0 \ .
\end{eqnarray}
We remark that ({\ref{kse2}}) is implied by the other KSE and
the bosonic conditions, however we shall retain ({\ref{kse2}})
for convenience.

\newsection{Reduction to Seven Dimensions}

The existence of the isometry generated by $h$ allows one
to reduce the KSE (and bosonic field equations)
to those on a 7-dimensional manifold $M_7$, where
\begin{eqnarray}
ds^2({\cal{S}}) = Q^{-2} h \otimes h + ds^2(M_7)
\end{eqnarray}
and we have set
\begin{eqnarray}
h^2 = Q^2
\end{eqnarray}
for non-zero constant $Q${\footnote{We remark that in \cite{hhor1},
the constant $Q$ was referred to as $k$; however in this work we shall
instead reserve $k$ to denote a parameter in the Aloff-Wallach space 
considered later in Section 5.}}. We shall set
\begin{eqnarray}
h = Q {\bf{e}}^8
\end{eqnarray}
and take frame indices $A, B=1, \dots , 7$ to be frame
indices on $M_7$. Then from the results of \cite{hhor1},
\begin{eqnarray}
{\tilde{H}} = Q^{-2} h \wedge dh + {\tilde{H}}_{(7)}
\end{eqnarray}
where ${\tilde{H}}_{(7)}$ is a 3-form on $M_7$, given in terms
of the $G_2$ 3-form,
$\varphi$, by
\begin{eqnarray}
\label{hhexp}
{\tilde{H}}_{(7)} = Q \varphi +e^{2 \Phi} \star_7 d \big(e^{-2 \Phi} \varphi \big) \ .
\end{eqnarray}
The $G_2$ 3-form,
$\varphi$, has components given by
\begin{eqnarray}
\parallel \eta_+ \parallel^2 \varphi_{ABC}= \langle \eta_+, \Gamma_8 \Gamma_{ABC} \eta_+ \rangle
\end{eqnarray}
and
\begin{eqnarray}
\parallel \eta_+ \parallel^2 \star_7 \varphi_{ABCD}=
\langle \eta_+, \Gamma_{ABCD} \eta_+ \rangle \ .
\end{eqnarray}

The Bianchi identity
$d {\tilde{H}}=0$ then implies
\begin{eqnarray}
\label{redbian}
d {\tilde{H}}_{(7)} = -Q^{-2} dh \wedge dh
\end{eqnarray}
and $dh$ is a closed 2-form on $M_7$, with
\begin{eqnarray}
dh \in {\mathfrak{g}}_2
\label{g2dh}
\end{eqnarray}
as a consequence of the supersymmetry. In addition, the
$G_2$ structure must be conformally co-calibrated
\begin{eqnarray}
\label{ccc}
d \big(e^{-2 \Phi} \star_7 \varphi \big)=0
\end{eqnarray}
which implies that
\begin{eqnarray}
\label{dilasolv}
{\hat{\nabla}}_A \Phi = {1 \over 12} {\varphi_A}{}^{B_1 B_2}{\hat{\nabla}}^D \varphi_{D B_1 B_2} \ ,
\end{eqnarray}
where ${\hat{\nabla}}$ denotes the Levi-Civita connection on $M_7$.
The gauge field equation ({\ref{gaugeeqn}}) is
\begin{eqnarray}
\label{Gauge7D}
d \big(e^{-2 \Phi} \star_7 {\tilde{H}}_{(7)} \big) =0~.
\end{eqnarray}
which is satisfied as a consequence of ({\ref{ccc}) and ({\ref{hhexp}}). The dilaton equation ({\ref{deqn}}) is equivalent to
\begin{eqnarray}
\label{redeq1}
{\hat{\nabla}}^A {\hat{\nabla}}_A \Phi -2 {\hat{\nabla}}^A \Phi {\hat{\nabla}}_A \Phi +{1 \over 4} Q^{-2}
dh_{AB} dh^{AB} \nonumber \\ \qquad \qquad +{1 \over 12} ({\tilde{H}}_{(7)})_{ABC} ({\tilde{H}}_{(7)})^{ABC} -{1 \over 2} Q^2 =0
\end{eqnarray}
and ({\ref{eineq}}) implies that
\begin{eqnarray}
\label{redeq2}
{\hat{R}}_{AB} = {1 \over 4} ({\tilde{H}}_{(7)})_{AMN}({\tilde{H}}_{(7)})_B{}^{MN} + Q^{-2} dh^M{}_A dh_{MB}
-2 {\hat{\nabla}}_A {\hat{\nabla}}_B \Phi
\end{eqnarray}
where ${\hat{R}}_{AB}$ denotes the Ricci tensor of $M_7$.
These conditions imply that
\begin{eqnarray}
\label{redeq3}
{\hat{R}} = {5 \over 12} ({\tilde{H}}_{(7)})_{ABC} ({\tilde{H}}_{(7)})^{ABC} +{3 \over 2} Q^{-2} dh_{AB} dh^{AB}
-4 {\hat{\nabla}}^A \Phi {\hat{\nabla}}_A \Phi -Q^2 \ .
\end{eqnarray}

The constant $Q$ is not free; it is determined by the $G_2$ structure via
\begin{eqnarray}
\label{kfix}
Q = -{1 \over 144} \star_7 \varphi^{ABCD} d \varphi_{ABCD}
\end{eqnarray}
as a consequence of the dilatino KSE. In particular, for nontrivial solutions with $Q \neq 0$, one cannot have $d \varphi=0$.

We remark that these conditions can be used to simplify certain components
of the Bianchi identity ({\ref{redbian}}). In particular, $dh \in {\mathfrak{g}}_2$
implies that $dh^{{\bf{7}}}=0$, and hence $(dh \wedge dh)^{{\bf{7}}}=0$. This, when combined with ({\ref{redbian}}), implies that the ${\bf{7}}$ part of $d {\tilde{H}}_{(7)}$ must also vanish, or equivalently
\begin{eqnarray}
(d {\tilde{H}}_{(7)})_{A B_1 B_2 B_3} \varphi^{B_1 B_2 B_3}=0 \ .
\end{eqnarray}
This condition can be rewritten as
\begin{eqnarray}
\label{bian7}
36Q {\hat{\nabla}}_A \Phi +6 {\hat{\nabla}}^N \Phi \varphi_N{}^{B_1 B_2} (\star_7 d \varphi)_{A B_1 B_2}
\nonumber \\
+\star_7 \varphi_A{}^{B_1 B_2 B_3} \big(-2 {\hat{\nabla}}^L \Phi {\hat{\nabla}}_L \varphi_{B_1 B_2 B_3}
+ {\hat{\nabla}}^L {\hat{\nabla}}_L \varphi_{B_1 B_2 B_3} \big)=0 \ .
\end{eqnarray}
Furthermore, it is possible to see that  ({\ref{bian7}}) also follows as a consequence of integrability conditions associated with the decomposition of $d \varphi$ in terms of torsion classes. In particular, on taking the exterior derivative of ({\ref{structure1}}), dualizing the
resulting 5-form, and then taking the ${\bf{7}}$ projection of this expression, one obtains
after some computation ({\ref{bian7}}).

Next, we consider the reduction of the KSE. First, note that
the $i=8$ component of ({\ref{kse1}}) can be rewritten,
using ({\ref{kse3}}) as
\begin{eqnarray}
\nabla_8 \eta_+ =0 \ .
\end{eqnarray}
Furthermore, with ({\ref{kse2}}), this is equivalent to
\begin{eqnarray}
\label{kos}
{\cal{L}}_h \eta_+ =0
\end{eqnarray}
where ${\cal{L}}$ here denotes the spinorial Lie derivative. The remaining content
of ({\ref{kse1}}), ({\ref{kse2}}) and ({\ref{kse3}}) is equivalent to
\begin{eqnarray}
\label{kse2a}
{\hat{\nabla}}_A \eta_+ + \bigg(-{1 \over 16} \Gamma_A (2 {\hat{\nabla}}_B \Phi \Gamma^B +Q \Gamma_8)
+{1 \over 96} \Gamma_A{}^{B_1 B_2 B_3} ({\tilde{H}}_{(7)})_{B_1 B_2 B_3} 
\nonumber \\
\qquad ~~~ -{3 \over 32}
({\tilde{H}}_{(7)})_{A B_1 B_2} \Gamma^{B_1 B_2} \bigg) \eta_+ =0
\nonumber \\
\end{eqnarray}
and
\begin{eqnarray}
\label{kse2b}
\big( 2 {\hat{\nabla}}_B \Phi \Gamma^B +Q \Gamma_8 \big)\eta_+ -{1 \over 6} ({\tilde{H}}_{(7)})_{ABC}\Gamma^{ABC} \eta_+ 
=0
\end{eqnarray}
and
\begin{eqnarray}
\label{kse2c}
dh_{AB} \Gamma^{AB} \eta_+ =0 \ .
\end{eqnarray}
The form of ({\ref{kse2a}}) is somewhat arbitrary, in the sense that one can
without loss of generality add any multiple of $\Gamma_A \times$({\ref{kse2b}}) to
({\ref{kse2a}}) and obtain an equivalent set of KSEs. However, the particular
form of ({\ref{kse2a}}) is inspired by the standard embedding
of heterotic solutions into IIB theory, as it is known how to formulate a Lichnerowicz type theorem in IIB \cite{iibhor, Gran:2013wca}. We next consider the heterotic Lichnerowicz type theorem,
which is derived using the same techniques utilized for the IIB horizons.

\subsection{Reduced Lichnerowicz Theorems}

We shall show that there exist near-horizon Dirac operators ${\cal{D}}^{(\pm)}$
with the property that ${\cal{D}}^{(\pm)} \eta_\pm =0$ if and only if 
$\eta_\pm$ satisfies ({\ref{kse2a}}), ({\ref{kse2b}}) and ({\ref{kse2c}}).
To proceed we define
\begin{eqnarray}
\label{kse3a}
{\hat{\nabla}}^{(\pm)}_A \eta_\pm &\equiv& 
{\hat{\nabla}}_A \eta_\pm + \bigg(-{1 \over 16} \Gamma_A (2 {\hat{\nabla}}_B \Phi \Gamma^B \pm Q \Gamma_8)
\nonumber \\
&+&{1 \over 96} \Gamma_A{}^{B_1 B_2 B_3} ({\tilde{H}}_{(7)})_{B_1 B_2 B_3} -{3 \over 32}
({\tilde{H}}_{(7)})_{A B_1 B_2} \Gamma^{B_1 B_2} \bigg) \eta_\pm 
\nonumber \\
\end{eqnarray}
and
\begin{eqnarray}
\label{kse3b}
{\cal{B}}^{(\pm)} \eta_\pm \equiv \big( 2 {\hat{\nabla}}_B \Phi \Gamma^B \pm Q \Gamma_8 \big)\eta_\pm -{1 \over 6} ({\tilde{H}}_{(7)})_{ABC}\Gamma^{ABC} \eta_\pm \ .
\end{eqnarray}
The reduced horizon Dirac operators associated with ({\ref{kse3a}}) are
\begin{eqnarray}
\label{reddir}
{\cal{D}}^{(\pm)} \eta_\pm \equiv \Gamma^A {\hat{\nabla}}_A \eta_\pm + \bigg(-{1 \over 24} ({\tilde{H}}_{(7)})_{ABC} \Gamma^{ABC}
-\Gamma^A {\hat{\nabla}}_A \Phi \mp{Q \over 2}\Gamma_8 \bigg) \eta_\pm \ .
\end{eqnarray}
Note that ${\hat{\nabla}}^{(+)}$ is the supercovariant derivative appearing in
({\ref{kse2a}}), and ${\cal{D}}^{(+)}$ is its associated reduced horizon Dirac equation.
Suppose that $\eta_\pm$ satisfies
\begin{eqnarray}
{\cal{D}}^{(\pm)} \eta_\pm =0 \ .
\end{eqnarray}
Then using ({\ref{redeq1}}), ({\ref{redeq2}}) and ({\ref{redeq3}}) and ({\ref{redbian}}), it is straightforward
to show, after some computation, that
\begin{eqnarray}
{\hat{\nabla}}^A {\hat{\nabla}}_A \parallel \eta_\pm \parallel^2
-2 {\hat{\nabla}}^A \Phi {\hat{\nabla}}_A \parallel \eta_\pm \parallel^2
&=&2 \langle {\hat{\nabla}}^{(\pm)A} \eta_\pm , {\hat{\nabla}}^{(\pm)}_A \eta_\pm \rangle
\nonumber \\
&+&{1 \over 8} Q^{-2} \parallel dh_{AB} \Gamma^{AB} \eta_\pm \parallel^2
\nonumber \\
&+&{9 \over 128} \parallel {\cal{B}}^{(\pm)} \eta_\pm \parallel^2 \ .
\nonumber \\
\end{eqnarray}
On applying the maximum principle, one obtains
\begin{eqnarray}
\parallel \eta_\pm \parallel^2= {\rm const.}
\end{eqnarray}
and
\begin{eqnarray}
\label{parallellichner}
{\hat{\nabla}}^{(\pm)}_A \eta_\pm =0, \qquad dh_{AB} \Gamma^{AB} \eta_\pm =0, \qquad {\cal{B}}^{(\pm)} \eta_\pm =0 \ .
\end{eqnarray}

It follows that if $\eta_\pm$ satisfies the reduced horizon
Dirac equation ${\cal{D}}^{(\pm)} \eta_\pm=0$ given by ({\ref{reddir}}), then
$\eta_\pm$ satisfies ({\ref{parallellichner}}).
Conversely, it is straightforward to show that if $\eta_\pm$ satisfies ({\ref{parallellichner}}) then $\eta_\pm$ satisfies the reduced horizon
Dirac equation ${\cal{D}}^{(\pm)} \eta_\pm=0$ given by ({\ref{reddir}}).

\newsection{A Condition for Supersymmetry Enhancement }

In this section we consider the necessary and sufficient conditions for supersymmetry enhancement from $N=2$ 
to $N \geq 4$. To establish the simplest form for such a condition,
we shall utilize the Lichnerowicz type theorem which holds for the Killing spinor
equation when reduced to $M_7$. From \cite{hhor1}, it is known that
if a $N=2$ solution described by spinors $\{ \eta_+, \eta_- \}$ admits supersymmetry enhancement to $N\geq4$, then extra spinors are given
by $\{ \Gamma_8 V_A \Gamma^A \eta_+, \Gamma_8 V_A \Gamma^A \eta_- \}$, where
$V$ is a certain vector field on $M_7$.

We aim to obtain the minimal set of conditions on such a $V$ which are necessary and sufficient to impose supersymmetry enhancement from $N=2$ to $N\geq 4$. In particular, suppose that $\eta_+$ is a solution of the ``+" chirality KSE ({\ref{kse1}}), ({\ref{kse2}}) and ({\ref{kse3}}). Define
\begin{eqnarray}
\eta'_+ = \Gamma_8 V_A \Gamma^A \eta_+ \ .
\end{eqnarray}
By construction, $\eta_+, \eta'_+$ are linearly independent as they are orthogonal
with respect to $\langle~,~ \rangle$.
Now consider the KSE. The Lichnerowicz type theorems established previously imply that it suffices to consider  ${\cal{D}}^{(+)} \eta'_+$, which is given by
\begin{eqnarray}
{\cal{D}}^{(+)} \eta'_+ &=& \Gamma_8 \bigg(-{1 \over 2} dV_{AB} \Gamma^{AB} + Q V_B \Gamma^B \Gamma_8 \bigg) \eta_+
\nonumber \\
&+& \Gamma_8 \bigg(-{\hat{\nabla}}^A V_A +2 V^A {\hat{\nabla}}_A \Phi \bigg) \eta_+ \ .
\end{eqnarray}
Hence, there is supersymmetry enhancement if and only if
\begin{eqnarray}
\label{instanton2}
\bigg({1 \over 2} dV_{AB} \Gamma^{AB} \Gamma_8 + Q  V_B \Gamma^B \bigg) \eta_+=0
\end{eqnarray}
and
\begin{eqnarray}
\label{nbauxa}
d \star_7 (e^{-2 \Phi} V)=0 \ .
\end{eqnarray}
The condition ({\ref{instanton2}}) is equivalent to
\begin{eqnarray}
\label{nbauxb}
Q V_A -{1 \over 2} \varphi_A{}^{B_1 B_2} dV_{B_1 B_2}=0
\end{eqnarray}
or equivalently
\begin{eqnarray}
\label{extrasusy}
dV-{1 \over 3}Q i_V \varphi \in {\mathfrak{g}}_2 \ .
\end{eqnarray}
Moreover, the condition ({\ref{ccc}}), together with ({\ref{nbauxb}}), implies ({\ref{nbauxa}}).
Hence, the necessary and sufficient condition for there to be supersymmetry enhancement is that there exists a
vector field $V$ on $M_7$ which satisfies ({\ref{extrasusy}}).

\newsection{An Exactly $N=2$ Heterotic Near-Horizon Solution}

In this section, we shall construct explicitly an example of a heterotic near-horizon solution
which satisfies the conditions for $N=2$ supersymmetry, but for which the supersymmetry
enhancement condition ({\ref{extrasusy}}) does not hold. This is therefore the first
known example of a heterotic near-horizon geometry preserving {\it exactly} $N=2$ supersymmetry.

Before presenting the solution, we remark that a number
of possible Riemannian manifolds equipped with $G_2$ structures are incompatible with
the conditions required for $N=2$ supersymmetry. For example, the $G_2$ structures constructed
in \cite{fernandez} and \cite{ivanov} satisfy $\varphi \wedge d \varphi=0$, and hence
$Q=0$ as a consequence of ({\ref{kfix}}). We therefore discard this case. Alternatively, we may consider co-calibrated solutions, for which $d \varphi = \lambda \star_7 \varphi$. Then ({\ref{kfix}}) implies that $\lambda =-{6 \over 7}Q$. Furthermore, the conformal co-calibration condition ({\ref{ccc}}) implies that $\Phi =const.$. It follows that
\begin{eqnarray}
{\tilde{H}}_{(7)} = {Q \over 7} \varphi
\end{eqnarray}
and hence the Bianchi identity ({\ref{redbian}})
implies
\begin{eqnarray}
{6 \over 49} Q^4 \star_7 \varphi = dh \wedge dh \ .
\end{eqnarray}
However, this condition can never hold, because in seven dimensions, there must be a non-vanishing vector field $Z$ such that $i_Z dh=0$, which would imply that $i_Z \star_7 \varphi=0$, and hence $Z=0$. Hence, there are no co-calibrated solutions
which are consistent with the conditions required for $N=2$ supersymmmetry.
Having eliminated these as possible candidates for constructing a solution, we
shall present a $G_2$ structure which is compatible with the conditions
of $N=2$ supersymmetry. The solution is obtained from the $G_2$ structure 
considered in \cite{cabrera}, in which $M_7$ is taken to be a compact Aloff-Wallach space
$M_7=M_{k,\ell}=SU(3)/U(1)_{k, \ell}$, where $U(1)_{k,\ell}$ is the circle subgroup of $SU(3)$ given by
\begin{eqnarray}
U(1)_{k,\ell}:=\left\{
\begin{bmatrix}
e^{ik\theta}&0&0 \\
0&e^{i\ell\theta}&0\\
0&0&e^{-i(k+\ell)\theta}
\end{bmatrix}
\middle | ~k, \ell \in\mathbb{Z}~,~|k|+|\ell|\ne 0~,~~\theta\in\mathbb{R}~  \right\}~.
\nonumber \\
\end{eqnarray}
An example of such a structure was constructed
in \cite{Doubrov:2011ef}, however for the particular choice of parameters used in
that solution, the Bianchi identity ({\ref{redbian}}) fails to hold, for essentially
the same reason as in the consideration of co-calibrated structures above.

\subsection{$G_2$ Structure and the $N=2$ Heterotic Solution}

Let us consider $M_7=M_{1,0} = SU(3) / U(1)_{1,0}$, where $M_{1,0}$ is an Aloff-Wallach \cite{aloffwallach} space with $k=1$ and $\ell=0$. Then, as is shown in Appendix B, the equations of the $G_2$ structure on $M_7$ are given by 
\begin{eqnarray}
de^1&=&-e^{23}-\sqrt{3}e^{45}+e^{67}+e^5\wedge\chi~
\nonumber \\
de^2&=&e^{13}+\sqrt{3}e^{46}-e^{57}+e^{6}\wedge\chi~
\nonumber \\
de^3&=&-e^{12}+e^{56}-2e^7\wedge\chi~
\nonumber \\
de^4&=&\sqrt{3}e^{15}-\sqrt{3}e^{26}~
\nonumber \\
de^5&=&-\sqrt{3}e^{14}+e^{27}-e^{36}-e^{1}\wedge\chi~
\nonumber \\
de^6&=&-e^{17}+\sqrt{3}e^{24}+e^{35}-e^2\wedge\chi~
\nonumber \\
de^7&=&e^{16}-e^{25}+2e^3\wedge\chi~
\nonumber \\
d\chi&=&e^{15}+e^{26}-2e^{37}~
\label{structureM7}
\end{eqnarray}
where $\{e^1,e^2,\dots e^7\}$ is an orthonormal co-frame, namely
\begin{eqnarray}
ds^2(M_7)=(e^1)^2+(e^2)^2+(e^3)^2+(e^4)^2+(e^5)^2+(e^6)^2+(e^7)^2~
\label{dsM7}
\end{eqnarray}
and $\chi$ is a 1-form on $M_7$. The fundamental 3-form $\varphi$ of the $G_2$ structure is given by
\begin{eqnarray}
\varphi=e^{123}-e^{167}+e^{257}-e^{356}+e^{145}+e^{246}+e^{347}~
\label{phi}
\end{eqnarray}
and the dual of $\varphi$ reads
\begin{eqnarray}
\star_7\varphi=e^{4567}-e^{2345}+e^{1346}-e^{1247}+e^{2367}+e^{1357}+e^{1256}~
\label{star7phi}
\end{eqnarray}
where $\epsilon_{1234567}=+1$. Using ({\ref{structureM7}}) and ({\ref{star7phi}}), it is straightforward to show that 
\begin{eqnarray}
d\star_7\varphi=0~.
\label{dstarphi}
\end{eqnarray}
Comparing ({\ref{dstarphi}}) with ({\ref{ccc}}), it follows that the dilaton is constant, that is $d\Phi=0$. Moreover, using ({\ref{kfix}}), we compute
\begin{eqnarray}
Q=-4~.
\label{k}
\end{eqnarray}
The constancy of the dilaton implies that ({\ref{hhexp}}) simplifies to
\begin{eqnarray}
\widetilde{H}_{(7)}=Q\varphi+\star_7 d\varphi~.
\label{H7}
\end{eqnarray}
Using ({\ref{phi}}) and ({\ref{structureM7}}), we compute $\widetilde{H}_{(7)}$ via ({\ref{H7}}), obtaining
\begin{eqnarray}
\widetilde{H}_{(7)}=-e^{123}+e^{167}-e^{257}+e^{356}-\sqrt{3}e^{145}+\sqrt{3}e^{246}~.
\label{H7frame}
\end{eqnarray}
Furthermore, let us define $dh=Qd\chi$. Using ({\ref{k}}) and ({\ref{structureM7}}), it follows that
\begin{eqnarray}
dh=-4(e^{15}+e^{26}-2e^{37})~.
\label{dh}
\end{eqnarray}
It is straightforward to show that $dh$, given by ({\ref{dh}}), satisfies ({\ref{g2dh}}). Moreover, we have found that the Bianchi identities ({\ref{redbian}}) and the bosonic field equations ({\ref{Gauge7D}})-({\ref{redeq2}}) are fulfilled by the above configuration.\\
\indent
It is interesting to inquire in which class of \cite{fgclass} our solution lies within. In particular, we investigate which torsion classes vanish.  It was shown in \cite{fgclass}  that there are 16 distinct classes of $G_2$ manifolds, which can be described in terms of the irreducible representations of the covariant derivative of the $G_2$ fundamental 3-form $\varphi$. It is also possible to characterize each class in terms of the irreducible representations of $d\varphi$ and $d\star_7\varphi$ \cite{Bryant}. To be precise, for any $G_2$ structure on a 7-dimensional orientable manifold $M_7$, there exist unique differential forms $\tau_0\in\Lambda^0(M_7)$, $\tau_1\in\Lambda^1_{\textbf{7}}(M_7)$, $\tau_2\in\Lambda^2_{\textbf{14}}(M_7)$, $\tau_3\in\Lambda^3_{\textbf{27}}(M_7)$ such that
\begin{eqnarray}
\label{structure1}
d\varphi&=&\tau_0\star_7\varphi+3\tau_1\wedge\varphi+\star_7\tau_3
\end{eqnarray}
and
\begin{eqnarray}
\label{structure2}
d\star_7\varphi&=&4\tau_1\wedge\star_7\varphi+\star_7\tau_2~.
\end{eqnarray}
Notice that
\begin{eqnarray}
\tau_0=\frac{1}{7}\star_7(\varphi\wedge d\varphi)
\label{tau0}
\end{eqnarray}
and $\tau_1=\frac{1}{12}\theta^{(7)}_{\varphi}$, where $\theta^{(7)}_{\varphi}$ is the Lee form on $M_7$
\begin{eqnarray}
\theta^{(7)}_{\varphi}=\star_7(\varphi\wedge\star_7d\varphi)~.
\label{tau1}
\end{eqnarray}
The differential forms $\tau_0,\tau_1,\tau_2,\tau_3$ are called \textit{intrinsic torsion forms} of the $G_2$ structure. Since each of them can be zero or non-zero, there are $2^4=16$ distinct classes of $G_2$ structures, as set out in Table C1 in Appendix C.

To begin the analysis of the torsion classes for our solution, comparing ({\ref{dstarphi}}) with ({\ref{structure2}}), it follows immediately that $\tau_1=\tau_2=0$. Moreover, comparing ({\ref{kfix}}) with ({\ref{structure1}}), we get
\begin{eqnarray}
\tau_0=-\frac{6}{7}Q=\frac{24}{7}~
\end{eqnarray}
thus $\tau_0\ne 0$.
Furthermore, an explicit computation shows that $\tau_3=\star_7d\varphi-\tau_0\varphi$ is given by
\begin{eqnarray}
\tau_3=\frac{3}{7}(-e^{123}+e^{167}-e^{257}+e^{356})+(\frac{4}{7}-\sqrt{3})e^{145}+(\frac{4}{7}+\sqrt{3})e^{246}+\frac{4}{7}e^{347}
\nonumber \\
\end{eqnarray}
thus $\tau_3\ne 0$. Consulting Table 1 of Appendix C, it follows that our solution belongs to the $\mathcal{W}_1\oplus\mathcal{W}_3$ class (co-closed $G_2$ structure).\\
\indent
In the following, we shall prove that the solution we have found above preserves exactly $N=2$ supersymmetries. By contradiction, let us assume that it preserves $N \geq 4$ supersymmetries. Then, as we have shown in Section 4, this implies that there exists a non-zero vector $V$ on $M_7$ such that ({\ref{extrasusy}}) is satisfied. Moreover, since $\eta_+$ and $\eta^{'}_+=\Gamma_8\slashed{V}\eta_+$ are Killing spinors, then 
\begin{eqnarray}
\slashed{V}\slashed{dh}\eta_+=0~
\label{Vdh}
\end{eqnarray}
and
\begin{eqnarray}
\slashed{dh}\slashed{V}\eta_+=0~.
\label{dhV}
\end{eqnarray}
Taking the difference of ({\ref{Vdh}}) and ({\ref{dhV}}), we get
\begin{eqnarray}
i_Vdh=0~.
\label{ivdh}
\end{eqnarray}
Substituting ({\ref{dh}}) into ({\ref{ivdh}}), it follows that
\begin{eqnarray}
V=V_4 e^4~
\label{V4}
\end{eqnarray}
where in ({\ref{V4}}) we use the same symbol $V$ to denote the 1-form which is dual to the vector $V$. $V_4$ is a non-zero constant, since $\eta^{'}_+$ has constant norm. Substituting ({\ref{V4}}) into ({\ref{extrasusy}}) and using ({\ref{k}}), we obtain
\begin{eqnarray}
-4V_A-\frac{1}{2}V_4(de^4)_{B_1B_2}\phi_A^{~~B_1B_2}=0~.
\label{enhext}
\end{eqnarray}
Using ({\ref{structureM7}}) and ({\ref{phi}}), it is easy to check that $de^4\in\mathfrak{g}_2$, which in turn implies that $V_A=0$ by means of ({\ref{enhext}}). Thus, by assuming that our solution preserves $N=4$ supersymmetries, we have reached a contradiction; this means that our solution preserves exactly $N=2$ supersymmetries. \\
\indent
To conclude this section, let us make an additional remark. Using ({\ref{V4}}), ({\ref{H7frame}}) and ({\ref{structureM7}}), it can be shown that $V=e^4$ is covariantly constant with respect to the connection with torsion, that is
\begin{eqnarray}
\hat{\nabla}_AV_B=\frac{1}{2}V^C(\widetilde{H}_{(7)})_{CAB}~
\end{eqnarray}
which in turn implies that $\eta^{'}_+$ satisfies the minimal $N=4$ gravitino KSE
\begin{eqnarray}
\hat{\nabla}_A\eta^{'}_+=\frac{1}{8}(\widetilde{H}_{(7)})_{ABC}\Gamma^{BC}\eta^{'}_+~.
\end{eqnarray}
The failure of ({\ref{extrasusy}}) to be satisfied corresponds to the fact that $\eta^{'}_+$ does \textit{not} satisfy the $N=4$ dilatino KSE
\begin{eqnarray}
\big( 2 {\hat{\nabla}}_B \Phi \Gamma^B +Q \Gamma_8 \big)\eta^{'}_+ -{1 \over 6} ({\tilde{H}}_{(7)})_{ABC}\Gamma^{ABC} \eta^{'}_+=0~. 
\end{eqnarray}
Such ``descendant" solutions, for which the gravitino equation holds for all the spinors, but the dilatino equation does not hold for all of the spinors satisfying the gravitino equation,
have also been considered in \cite{Papadopoulos:2009br} and \cite{Gran:2007fu}.

\newsection{Conclusion}

We have found that there exists a near-horizon solution of heterotic supergravity
preserving exactly $N=2$ supersymmetry, utilizing the family of $G_2$ structures
constructed in \cite{cabrera}. Although many supersymmetric heterotic near-horizon
geometries had previously been found \cite{hhor2}, these solutions all preserved at least
$N=4$ supersymmetry. The solution found in this paper is the first near-horizon geometry
to preserve the minimal $N=2$ supersymmetry. This demonstrates that there is not some 
additional mechanism for supersymmetry enhancement of near-horizon solutions in the heterotic theory, which would have meant that the minimal amount of supersymmetry preserved would be $N=4$ and not $N=2$. 

We remark that this $N=2$ solution has constant dilaton. Indeed, all known supersymmetric heterotic near-horizon solutions have constant dilaton. It would be interesting to determine whether or not this is a generic property, analogous to an attractor mechanism argument. The constancy of scalars for supersymmetric near-horizon geometries holds for some theories, such as ungauged $N=2$, $D=4$ supergravity coupled to $U(1)$ vector multiplets \cite{Gutowski:2016gkg}, but not for others, e.g. gauged $N=2$, $D=4$ supergravity coupled to $U(1)$ vector multiplets \cite{Klemm:2010mc, Klemm:2011xw,
Gnecchi:2013mja, Gutowski:2016gkg}.
 It is therefore not a priori apparent whether or not one might expect the heterotic dilaton $\Phi$ to be generically constant in the near-horizon limit. However, it would be interesting to determine if the ${\bf{7}}$ component of the Bianchi identity, given in ({\ref{bian7}}), can be used to obtain additional conditions on the dilaton.

\vskip 0.5cm
\noindent{\bf Acknowledgements} \vskip 0.1cm
\noindent  DF is partially supported by the STFC DTP Grant
ST/S505742. JG is supported by the STFC Consolidated Grant ST/L000490/1.
JG would like to thank Prof. Ulf Gran for hospitality during a 2018 visit
to the Department of Physics, Chalmers University of Technology, at which
part of this work was done.

\vskip 0.5cm

\noindent{\bf Data Management:} \vskip 0.1cm
\noindent  No additional research data beyond the data presented and cited in this work are needed to validate the research findings in this work.
\vskip 0.5cm

\setcounter{section}{0}
\setcounter{subsection}{0}
\setcounter{equation}{0}

\appendix{Useful $G_2$ identities}

A seven dimensional orientable Riemannian manifold $M_7$ with a
$G_2$ structure admits a 3-form $\varphi$, with Hodge dual $\star_7 \varphi$.
These forms satisfy several algebraic identities;
\begin{eqnarray}
\varphi_{ABJ} \varphi^{CDJ} = 2 \delta^{CD}_{AB}-\star_7\varphi_{AB}{}^{CD}
\end{eqnarray}
and hence
\begin{eqnarray}
\varphi_{ACD} \varphi^{BCD} = 6 \delta_A^B
\end{eqnarray}
and
\begin{eqnarray}
\varphi_{ABJ} \star_7\varphi^{CDLJ} =6 \delta_{[A}^{[C} {\varphi^{D L]}}_{B]} \ .
\end{eqnarray}
where $A,B= 1, \dots, 7$. In addition, we have
\begin{eqnarray}
\epsilon_{A_1 A_2 A_3}{}^{B_1 B_2 B_3 B_4} &=& - \varphi_{[A_1 A_2}{}^{[B_1} \star_7\varphi_{A_3]}{}^{B_2 B_3 B_4]} +3 \varphi^{[B_1 B_2}{}_{[A_1} \star_7\varphi_{A_2 A_3]}{}^{B_3 B_4]}
\nonumber \\
&-& \varphi^{[B_1 B_2 B_3} \star_7\varphi_{A_1 A_2 A_3}{}^{B_4]}
\end{eqnarray}
and
\begin{eqnarray}
\star_7\varphi_{A_1 A_2 A_3 C} \star_7 \varphi^{B_1 B_2 B_3 C}
&=& 6 \delta_{A_1 A_2 A_3}^{B_1 B_2 B_3}
-9 \delta_{[A_1}^{[B_1} \star_7\varphi_{A_2 A_3]}{}^{B_2 B_3]}
\nonumber \\
&-&\varphi_{A_1 A_2 A_3} \varphi^{B_1 B_2 B_3}
\end{eqnarray}
and
\begin{eqnarray}
\varphi_{A_1 A_2 A_3} \star_7\varphi^{B_1 B_2 B_3 B_4}
&=& 5 \varphi_{[A_1 A_2}{}^{[B_1} \star_7\varphi_{A_3]}{}^{B_2 B_3 B_4]}
\nonumber \\
&+&3 \varphi^{[B_1 B_2}{}_{[A_1} \star_7\varphi_{A_2 A_3]}{}^{B_3 B_4]}
\nonumber \\
&-&3 \varphi^{[B_1 B_2 B_3} \star_7\varphi_{A_1 A_2 A_3}{}^{B_4]}
\end{eqnarray}
and
\begin{eqnarray}
\varphi_{[A_1 A_2}{}^{[B_1} \star_7\varphi_{A_3]}{}^{B_2 B_3 B_4]}
= \varphi^{[B_1 B_2 B_3} \star_7\varphi_{A_1 A_2 A_3}{}^{B_4]} 
-6 \delta_{[A_1}^{[B_1} \delta_{A_2}^{B_2} \varphi^{B_3 B_4]}{}_{A_3]} \ .
\nonumber \\
\end{eqnarray}

The $q$-forms on $M_7$ decompose w.r.t. various
irreps of $G_2$

\begin{eqnarray}
\Lambda^1 &=& \Lambda^1_{\bf{7}}
\nonumber \\
\Lambda^2 &=& \Lambda^2_{\bf{7}} \oplus \Lambda^2_{\bf{14}}
\nonumber \\
\Lambda^3 &=& \Lambda^3_{\bf{1}} \oplus \Lambda^3_{\bf{7}} \oplus \Lambda^3_{\bf{27}}
\nonumber \\
\Lambda^4 &=& \Lambda^4_{\bf{1}} \oplus \Lambda^4_{\bf{7}} \oplus \Lambda^4_{\bf{27}}
\nonumber \\
\Lambda^5 &=& \Lambda^5_{\bf{7}} \oplus \Lambda^5_{\bf{14}}
\nonumber \\
\Lambda^6 &=& \Lambda^6_{\bf{7}} \ .
\end{eqnarray}
For our purposes, the projections associated with the 2-forms, 3-forms and 4-forms are of most interest, and
we find that
\begin{eqnarray}
(P^{\bf{7}} \alpha)_{A_1 A_2} &=& {1 \over 3} \alpha_{A_1 A_2} -{1 \over 6} (\star_7\varphi)_{A_1 A_2}{}^{B_1 B_2}
\alpha_{B_1 B_2}
\nonumber \\
(P^{\bf{14}} \alpha)_{A_1 A_2} &=& {2 \over 3} \alpha_{A_1 A_2} +{1 \over 6} (\star_7\varphi)_{A_1 A_2}{}^{B_1 B_2}
\alpha_{B_1 B_2}
\end{eqnarray}
where $\alpha$ is a 2-form. In particular, $\alpha \in {\mathfrak{g}}_2$ iff $\alpha^{\bf{7}}=0$, which is equivalent to the condition
\begin{eqnarray}
\varphi_A{}^{BC} \alpha_{BC}=0 \ .
\end{eqnarray}
For the 3 forms,
\begin{eqnarray}
(P^{\bf{1}} \alpha)_{A_1 A_2 A_3} &=& {1 \over 42} \varphi^{B_1 B_2 B_3} \alpha_{B_1 B_2 B_3}
\varphi_{A_1 A_2 A_3}
\nonumber \\
(P^{\bf{7}} \alpha)_{A_1 A_2 A_3} &=& {1 \over 4} \alpha_{A_1 A_2 A_3}-{1
\over 24} \varphi^{B_1 B_2 B_3} \alpha_{B_1 B_2 B_3}
\varphi_{A_1 A_2 A_3} 
\nonumber \\
&-&{3 \over 8} \alpha_{B_1 B_2 [A_1} \star_7\varphi_{A_2 A_3]}{}^{B_1 B_2}
\nonumber \\
(P^{\bf{27}} \alpha)_{A_1 A_2 A_3} &=& {3 \over 4} \alpha_{A_1 A_2 A_3}+{1
\over 56} \varphi^{B_1 B_2 B_3} \alpha_{B_1 B_2 B_3}
\varphi_{A_1 A_2 A_3}
\nonumber \\
 &+&{3 \over 8} \alpha_{B_1 B_2 [A_1} \star_7\varphi_{A_2 A_3]}{}^{B_1 B_2}
\nonumber \\
\end{eqnarray}
where $\alpha$ is a 3-form; and for the 4 forms
\begin{eqnarray}
(P^{\bf{1}} \alpha)_{A_1 A_2 A_3 A_4} &=&{1 \over 168} \alpha^{B_1 B_2 B_3 B_4}
(\star_7\varphi)_{B_1 B_2 B_3 B_4} (\star_7\varphi)_{A_1 A_2 A_3 A_4}
\nonumber \\
(P^{\bf{7}} \alpha)_{A_1 A_2 A_3 A_4} &=&{1 \over 4} \alpha_{A_1 A_2 A_3 A_4}-{3 \over 4} (\star_7\varphi)^{B_1 B_2}{}_{[A_1 A_2} \alpha_{A_3 A_4] B_1 B_2}
\nonumber \\
&-&{1 \over 96} \alpha^{B_1 B_2 B_3 B_4}
(\star_7\varphi)_{B_1 B_2 B_3 B_4} (\star_7\varphi)_{A_1 A_2 A_3 A_4}
\nonumber \\
(P^{\bf{27}} \alpha)_{A_1 A_2 A_3 A_4} &=&{3 \over 4} \alpha_{A_1 A_2 A_3 A_4}+{3 \over 4} (\star_7\varphi)^{B_1 B_2}{}_{[A_1 A_2} \alpha_{A_3 A_4] B_1 B_2}
\nonumber \\
&+&{1 \over 224} \alpha^{B_1 B_2 B_3 B_4}
(\star_7\varphi)_{B_1 B_2 B_3 B_4} (\star_7\varphi)_{A_1 A_2 A_3 A_4}
\nonumber \\
\end{eqnarray}
where $\alpha$ is a 4-form. In particular, for a 4-form $\alpha$,
$\alpha^{\bf{7}}=0$ if and only if
\begin{eqnarray}
\alpha_{ABCL} \varphi^{BCL}=0 \ .
\end{eqnarray}

\appendix{Derivation of the $G_2$ structure}

In this Appendix, we present further details of how the $G_2$ structure presented in Section 5.1
is derived, following \cite{cabrera}. The structure equations on the 8-dimensional horizon $\mathcal{S}$ are given by
\begin{eqnarray}
df^1&=&-f^{23}-\sqrt{3}f^{45}+f^{67}+f^{58}
\nonumber \\
df^2&=&f^{13}+\sqrt{3}f^{46}-f^{57}+f^{68}
\nonumber \\
df^3&=&-f^{12}+f^{56}-2f^{78}
\nonumber \\
df^4&=&\sqrt{3}f^{15}-\sqrt{3}f^{26}
\nonumber \\
df^5&=&-\sqrt{3}f^{14}+f^{27}-f^{36}-f^{18}
\nonumber \\
df^6&=&-f^{17}+\sqrt{3}f^{24}+f^{35}-f^{28}
\nonumber \\
df^7&=&f^{16}-f^{25}+2f^{38}
\nonumber \\
df^8&=&f^{15}+f^{26}-2f^{37}.
\label{structure8}
\end{eqnarray}
In ({\ref{structure8}}), to simplify the solution we construct, we have set the parameters of the $G_2$ structure in \cite{cabrera} to the following values
\begin{eqnarray}
a=b=c=d=\frac{1}{\sqrt{2}}~,~~~~k=1~,~~~\ell=0~.
\end{eqnarray}

The metric on $\mathcal{S}$ is given by
\begin{eqnarray}
ds^2(\mathcal{S})=(f^8)^2+ds^2(M_7)~
\end{eqnarray}
where 
\begin{eqnarray}
ds^2(M_7)=(f^1)^2+(f^2)^2+(f^3)^2+(f^4)^2+(f^5)^2+(f^6)^2+(f^7)^2~
\label{dsM72}
\end{eqnarray}
and the fundamental 3-form $\varphi$ of the $G_2$ structure is given by
\begin{eqnarray}
\varphi=f^{123}-f^{167}+f^{257}-f^{356}+f^{145}+f^{246}+f^{357}~.
\label{phi2}
\end{eqnarray}
In order to reduce the structure equations ({\ref{structure8}}) down to $M_7$, consider the frame transformation
\begin{eqnarray}
f^A\to e^A=X^A_{~~B}f^B~
\label{fea}
\end{eqnarray}
where $A,B=1,2,\dots 7$ and $X \in SO(7)$. Enforcing the requirement
\begin{eqnarray}
\mathcal{L}_8 e^A=0
\end{eqnarray}
amounts to imposing the differential equation
\begin{eqnarray}
\frac{\partial}{\partial\tau}X^A_{~~B}-X^A_{~~C}C^C_{~~B8}=0~
\label{Xdiff}
\end{eqnarray}
where the constants $C^C_{~~B8}$ are defined by ($i,j=A,8$)
\begin{eqnarray}
df^i=\frac{1}{2}C^i_{~jk}f^j\wedge f^k~
\end{eqnarray}
and we have set $\partial_8=\frac{\partial}{\partial\tau}$, for some local coordinate $\tau$. Moreover, we take the 1-form dual to $\frac{\partial}{\partial\tau}$ to be
\begin{eqnarray}
f^8=d\tau+\chi~
\label{f8}
\end{eqnarray}
where $\chi$ is a 1-form on $M_7$. Notice that $df^8=d\chi$. The solution of ({\ref{Xdiff}}) is given by
\begin{eqnarray}
X^A_{~B}=
\begin{bmatrix}
\cos\tau&0&0&0&\sin\tau&0&0 \\
0&\cos\tau&0&0&0&\sin\tau&0\\
0&0&\cos(2\tau)&0&0&0&-\sin(2\tau)\\
0&0&0&1&0&0&0\\
-\sin\tau&0&0&0&\cos\tau&0&0 \\
0&-\sin\tau&0&0&0&\cos\tau&0\\
0&0&\sin(2\tau)&0&0&0&\cos(2\tau)
\label{X}
\end{bmatrix}
~.
\nonumber \\
\label{X}
\end{eqnarray}
Notice that $X=(X^A_{~~B})\in G_2$. In order to show that, rewrite ({\ref{Xdiff}}) as follows
\begin{eqnarray}
C^M_{~~B8}=(X^{-1})^M_{~~A}\frac{\partial}{\partial\tau}X^A_{~~B}
\label{XdX}
\end{eqnarray}
Using ({\ref{structure8}}), it can be easily checked that $\omega_{AB}:=C_{AB8}=\delta_{AC}C^C_{~~B8}$ is a 2-form on $M_7$ and $\omega_{AB}\varphi^{AB}_{~~~C}=0$, thus $\omega\in\mathfrak{g}_2$. Equation ({\ref{XdX}}) then implies that $X\in G_2$. In turn, this means that the 3-form $\varphi$, defined by ({\ref{phi2}}), is left unchanged by ({\ref{fea}}), that is
\begin{eqnarray}
\varphi=e^{123}-e^{167}+e^{257}-e^{356}+e^{145}+e^{246}+e^{357}
\label{phi3}
\end{eqnarray}
which coincides with ({\ref{phi}}). Eventually, using ({\ref{X}}), ({\ref{fea}}) and ({\ref{structure8}}), a tedious but straightforward computation yields ({\ref{structureM7}}).

\appendix{The 16 classes of $G_2$ manifolds}

The $16$ distinct classes of $G_2$ structures, determined in terms of their torsion
classes, are summarized in the following table \cite{fgclass}:
\begin{table}[h!]
\begin{center}
 \begin{tabular}{||c| c ||} 
 \hline
 Class & Defining relation  \\ [0.5ex] 
 \hline\hline
 $\mathcal{P}$ & $\tau_0=\tau_1=\tau_2=\tau_3=0 $\\ 
 \hline
 $\mathcal{W}_1 $&  $\tau_1=\tau_2=\tau_3=0 $\\
 \hline
$ \mathcal{W}_2$ &  $\tau_0=\tau_1=\tau_3=0$ \\
 \hline
$ \mathcal{W}_3$ &  $\tau_0=\tau_1=\tau_2=0 $\\
 \hline
$ \mathcal{W}_4$ & $ \tau_0=\tau_2=\tau_3=0$ \\
 \hline
  $\mathcal{W}_1\oplus\mathcal{W}_2$&  $\tau_1=\tau_3=0$ \\
 \hline
 $  \mathcal{W}_1\oplus\mathcal{W}_3$& $ \tau_1=\tau_2=0$ \\
 \hline
 $  \mathcal{W}_2\oplus\mathcal{W}_3$&$  \tau_0=\tau_1=0 $\\
 \hline
  $ \mathcal{W}_1\oplus\mathcal{W}_4$&$  \tau_2=\tau_3=0$ \\
 \hline
  $ \mathcal{W}_2\oplus\mathcal{W}_4$& $ \tau_0=\tau_3=0$ \\
 \hline
   $\mathcal{W}_3\oplus\mathcal{W}_4$& $ \tau_0=\tau_2=0$ \\
 \hline
   $\mathcal{W}_1\oplus\mathcal{W}_2\oplus\mathcal{W}_3$&$  \tau_1=0 $\\
 \hline
 $   \mathcal{W}_1\oplus\mathcal{W}_2\oplus\mathcal{W}_4$&$  \tau_3=0$ \\
 \hline
 $  \mathcal{W}_1\oplus\mathcal{W}_3\oplus\mathcal{W}_4$&$  \tau_2=0$ \\
 \hline
  $  \mathcal{W}_2\oplus\mathcal{W}_3\oplus\mathcal{W}_4$&  $\tau_0=0$ \\
 \hline
 $   \mathcal{W}$&  \textrm{no relation}\\ [1ex]
\hline
\end{tabular}
\caption{The 16 classes of $G_2$ manifolds in terms of the intrinsic torsion forms $\tau_i$.}
\end{center}
\label{tablec1}
\end{table}

\end{document}